\documentstyle[preprint,aps,floats]{revtex}
\tightenlines
\input psfig.tex 
\begin{document}
\def\ba{\begin{eqnarray}}
\def\ea{\end{eqnarray}}
\def\be{\begin{equation}}
\def\ee{\end{equation}}
\def\({\left(}
\def\){\right)}
\def\[{\left[}
\def\]{\right]}
\def\lagrange {{\cal L}}
\def\del {\nabla}
\def\d {\partial}
\def\Tr{{\rm Tr}}
\def\half{{1\over 2}}
\def\fourth{{1\over 8}}
\def\bibi{\bibitem}
\def\S{{\cal S}}
\def\H{{\cal H}}
\def\xx{\mbox{\boldmath $x$}}
\newcommand{\phpr} {\phi_0^{\prime}}
\newcommand{\gam}{\gamma_{ij}}
\newcommand{\sqgam}{\sqrt{\gamma}}
\newcommand{\dph}{\delta\phi}
\newcommand{\om} {\Omega}
\newcommand{\dom}{\delta^{(3)}\left(\Omega\right)}
\newcommand{\rar}{\rightarrow}
\newcommand{\Rar}{\Rightarrow}
\newcommand{\labeq}[1] {\label{eq:#1}}
\newcommand{\eqn}[1] {(\ref{eq:#1})}
\newcommand{\labfig}[1] {\label{fig:#1}}
\newcommand{\fig}[1] {\ref{fig:#1}}
\def\gsim{ \lower .75ex \hbox{$\sim$} \llap{\raise .27ex \hbox{$>$}} }
\def\lsim{ \lower .75ex \hbox{$\sim$} \llap{\raise .27ex \hbox{$<$}} }
\newcommand\bigdot[1] {\stackrel{\mbox{{\huge .}}}{#1}}
\newcommand\bigddot[1] {\stackrel{\mbox{{\huge ..}}}{#1}}

 \title{Gravity Waves from Instantons}

\author{Thomas Hertog\thanks{Aspirant FWO-Vlaanderen}\hspace{-.05in}
\thanks{email:T.Hertog@damtp.cam.ac.uk}
$\ $and 
Neil Turok\thanks{email:N.G.Turok@damtp.cam.ac.uk}\\
DAMTP, Silver St, Cambridge, CB3 9EW, U.K.}
\date{\today}
\maketitle

\begin{abstract}
We perform a first principles computation of the spectrum of gravity waves 
produced in open inflationary universes. The background spacetime 
is taken to
be the continuation of an 
instanton saddle point of the Euclidean no boundary path integral.
The two-point tensor correlator is computed directly from the
path integral and is shown to be unique and well behaved in the
infrared. 
We discuss the tensor contribution to
the cosmic microwave background anisotropy and show how
it
may provide an observational discriminant
between different types of primordial instantons. 
\end{abstract}
\vskip .2in

\section{Introduction}

The inflationary universe scenario provides an appealing explanation for the 
smoothness and flatness of the present universe,
as well as a mechanism for the
origin of density fluctuations. 
Until recently it was believed that inflation inevitably 
predicted a flat
$\Omega_0 =1$ universe. However in \cite{Bucher} 
it was shown that with mild fine tuning 
an open universe is also possible.
The potential must have 
a sharp false vacuum in which the field is assumed to have 
become trapped. The field is then assumed to 
tunnel out via an instanton known as the Coleman-De Luccia instanton
\cite{Coleman}, producing a bubble 
within which slow roll inflation occurs. 
The interior of the bubble produced
via the Coleman-De Luccia  instanton is an infinite inflating open universe.

Such models provide important counter-examples to the
standard folklore but require quite contrived scalar field 
potentials. Recently, however, Hawking and one of us showed that open
inflation can occur much more generally. 
We found 
a new class of instantons \cite{Hawking}
that exist for essentially any inflationary potential, and
provide saddle points of the 
Euclidean path integral. The continuation of these instantons
is similar to that of the Coleman-De Luccia instantons, and they
define initial conditions for open inflationary universes.
Although the Hawking-Turok instantons are singular, the singularity is mild
enough for
the quantization of perturbations to
be well posed \cite{Hawking},\cite{Gar},\cite{Gratton}. In this paper
we compute the spectrum of gravity waves for both Coleman-De Luccia
and Hawking-Turok instantons. 

This paper is a companion to ref. \cite{Gratton}, where
scalar fluctuations about open inflationary instantons 
were calculated. 
Here we perform the analogous calculation for the
tensor fluctuations and discuss possible observable signals
in the 
CMB anisotropy power spectrum.
The calculation  is performed in
the framework of the 
Euclidean no boundary proposal due to  Hartle and Hawking
\cite{Hartle}, as discussed in \cite{Gratton}. The 
correlator is computed in the 
Euclidean region where it is uniquely determined by
a Gaussian integral, and then analytically continued in 
the coordinates of the classical background solution 
into the Lorentzian region of interest.
Our main result for the 
tensor correlator (\ref{lorcorf}) is given in a form which
is straightforward to compute numerically. We defer detailed 
numerical calculations of the CMB anisotropies to a future paper
\cite{gratthertog} in which both scalar and tensor contributions
for a variety of scalar potentials will be discussed. 

There have over the last few years been many papers 
exploring similar calculations, mostly making 
one approximation or another  \cite{sasrefs}, \cite{Sasaki}, \cite{Tanaka}. 
Very recently, Garriga {\it et. al.}
have independently obtained 
formulae for the scalar and tensor correlators similar to ours
 \cite{Garriga}.
These formulae have been numerically implemented in 
ref. \cite{LST} which gives results for some examples of
Coleman-De Luccia instantons 
calculated without approximation. 

We feel that the derivation given here is significantly
clearer than in these papers, and that 
our method has several important conceptual
advantages. First, all earlier authors performed a mode by mode analysis.
In this framework, one requires a prescription for the vacuum state
for each perturbation mode 
and this is taken to be the state in which the positive frequency
modes are regular on the lower half of
the instanton.  This prescription is rather {\it ad hoc}. 
In contrast, our method is to simply perform the Euclidean
no boundary path integral. This automatically gives a unique Greens function.
There is no need for an additional prescription, indeed imposing 
one 
is contrary to the spirit of 
the no boundary proposal (see the discussion in \cite{Gratton}). The whole
idea of the Euclidean no boundary proposal 
is that an essentially {\it topological} prescription should define 
the initial state of the universe. Analyticity arises because the
background solution is a solution to a differential equation.
Divergent fluctuation modes have infinite Euclidean action and
are therefore suppressed in the path integral.
Second, in the matching method of 
Garriga {\it et. al.}, they devote a great deal of effort to determining
the action for perturbations in region II, the part of the Lorentzian spacetime
exterior to the open universe region. This introduces considerable
technical complexity since the spatial hypersurfaces used in their
canonical quantisation approach are inhomogeneous in region 
II.
Our approach is to analytically continue directly from the Euclidean 
region into the open universe. 
Region II is just a part of the continuation route with 
no special significance. Third, as emphasised in \cite{Gratton}, 
we deal throughout directly
with the real space correlator. In this  approach 
`super-curvature' modes are automatically included and their relation to the 
`sub-curvature' modes is thereby made clear. A related fact is
that 
we find that the real space correlator to be infrared finite
even in perfect de Sitter space, as mentioned below.
Finally, Garriga {\it et. al.} only give formulae for equal-time correlators.
To compute the microwave anisotropies one requires the unequal-time correlator,
which we give here. We are also careful to define the continuation of
the conformal time coordinate into the Euclidean region, which is
not explained in ref. \cite{LST}.

The paper is organised as follows.
In section 2 we describe the relevant path integral and
the model-dependent Schr\"{o}dinger  operator which occurs in the 
Euclidean action. We show that for singular instantons
the singularity acts as a reflecting 
boundary, fixing Dirichlet boundary conditions for the perturbation modes
\cite{Hawking},\cite{Gar},\cite{Gratton}.
The Euclidean tensor correlator is computed from the 
path integral in section 3.
In this calculation we need several 
properties 
of maximally symmetric bitensors on $S^3$, which are 
described in Appendix A.
Section 4 describes the analytic continuation to the 
open universe.
Finally, section 5 is devoted to the Sachs-Wolfe integral
to determine the contribution of gravitational waves to the CMB anisotropy.
Here we 
comment on 
possible observational 
distinctions between  
Coleman-De Luccia and Hawking-Turok 
instantons.

We conclude this introduction with two technical remarks.
First, the question of discrete `supercurvature' modes
arises in the tensor calculation just as in the scalar case
\cite{Gratton}. Here, however, we 
find that although the relevant Schr\"{o}dinger  operator
possesses a bound state just as in the case of scalar perturbations
\cite{Gratton}, here it
does not generate a `super-curvature mode'.
Instead the relevant mode is a time-independent 
shift in the metric perturbation
which may be gauged away. This is in agreement with refs. 
\cite{Tanaka},\cite{Garriga}. Second, 
it has been claimed in the much of the previous literature 
that the spectrum of gravity waves 
in pure de Sitter space
is infrared divergent \cite{allencald}, \cite{sasrefs}
but that the divergence 
disappears once the existence of the bubble
wall is taken in account \cite{Sasaki}. 
In our approach we find a different result. 
Neglecting the gauge mode previously mentioned, the two point correlator 
has a well defined long-wavelength limit
even in perfect de Sitter space. We shall investigate this issue further 
in future work.

\section{The Path Integral for Tensor Fluctuations}

In quantum cosmology the basic object is the wavefunctional
$\Psi \left[h_{ij},\phi \right]$, the amplitude for 
a three-geometry with metric
$h_{ij}$ and field configuration $\phi$. It is formally 
given by a path integral
\ba
\label{uni}
\Psi \left[h_{ij},\phi \right] \sim
\int^{h_{ij},\phi}\left[{\cal D} g\right] \left[{\cal D}\phi \right] e^{iS[ g,\phi ] }.
\ea
Following Hartle and Hawking 
\cite{Hartle} the lower limit of
the path integral is defined by 
continuing to Euclidean time and integrating over all compact 
Riemannian metrics $g$ and
field configurations $\phi$. If one can find a saddle point of (\ref{uni}),
namely a classical solution satisfying the Euclidean no boundary condition,
one can in principle at least compute the path integral 
as a perturbative expansion to any desired power in  $\hbar$.

In this paper, we shall compute
the two-point 
tensor fluctuation correlator about classical 
solutions describing the beginning of open inflationary universes, 
to first order in $\hbar$. The principles are described in
\cite{Gratton}, namely that we compute the correlator in
the Euclidean region where the exponent $iS$ in the path integral
becomes $-S_E=-(S_0+S_2)$, 
where $S_E$ is the Euclidean action, $S_0$ is the instanton
action and $S_2$ the action for fluctuations. We shall keep the latter
only to second order, this being all that is 
needed to compute the quantum fluctuations
to leading order in $\hbar$. The correlator is then given 
by a 
Gaussian path integral
\ba\label{wave}
\langle t_{ij}(x)t_{i'j'}(x')\rangle =
{ \int \left[{\cal D} \delta g\right]\left[{\cal D} \delta \phi\right]
e^{-S_2}t_{ij}(x)t_{i'j'}(x') \over  \int \left[{\cal D} \delta g\right]
\left[{\cal D} \delta \phi\right] 
e^{-S_2}}.
\ea
The  Lorentzian correlator is then obtained by
analytically continuing in the coordinates of the 
background classical solution, into the open inflating region. 

The $O(4)$ symmmetric instantons of interest possess a line
element of the form
\newline
$d\sigma^2+b^2(\sigma) d\Omega_3^2$ where $d\Omega_3^2$
is the line element on $S^3$. Both Hawking-Turok and Coleman-De Luccia 
instantons possess a regular pole which we take to be at $\sigma=0$. 
As $\sigma$ approaches zero, we have $b(\sigma) \rightarrow \sigma$. 
The  Coleman-De Luccia instantons have a second regular pole
where $b \rightarrow 
\sigma_m-\sigma$ where $\sigma_m$ is the maximum value of $\sigma$.
In contrast Hawking-Turok instantons
have $b \rightarrow  (\sigma_m-\sigma)^{1\over 3}$ as $\sigma \rightarrow \sigma_m$.
It is useful in both cases to introduce a 
conformal spatial coordinate satisfying $dX=d\sigma/b(\sigma)$, 
so that the line element takes the form 
\begin{eqnarray}
ds^2 & = & b^2 (X) \left( dX^2+ d\Omega_3^2\right)
\end{eqnarray}
For Hawking-Turok instantons we define 
\begin{eqnarray}
X &\equiv& \int_{\sigma}^{\sigma_{m}} \frac{d\sigma'}{b(\sigma')}
.
\end{eqnarray}
so $X=0$ corresponds to the singular pole and
$X \rightarrow \infty$ to the regular pole.
For Coleman-De Luccia instantons $X$
may be conveniently defined by $\int_{\sigma}^{\sigma_{t}} \frac{d\sigma'}{b(\sigma')}$,
where $\sigma_{t}$ is the value of sigma for which $b$ is a maximum,
and then $X$ ranges from $-\infty$ to $+\infty$.
We write the perturbed line element and the scalar field as
\begin{eqnarray}
ds^2 & = & b^2 (X) \left( ( 1+2A)dX^2 + S_{i}dx^{i} dX +
(\gamma_{ij} + h_{ij}) dx^{i}dx^{j} \right),\nonumber\\
\phi & = & \phi_0 (X) + \delta \phi.
\end{eqnarray}
and decompose $S_{i}$ and $h_{ij}$ as follows \cite{Kodama}
\begin{eqnarray}\label{dec}
h_{ij} & = & \frac{1}{3} h \gamma_{ij} + 2 \left(
\nabla _{i} \nabla_{j}  - \frac{\gamma_{ij}}{3} \Delta_{3}\right) E
+ 2 F_{(i \vert j)} + t_{ij},\nonumber\\
S_{i} & = & B_{\vert i} + V_{i}.
\end{eqnarray}
Here $\Delta_{3}$ is the Laplacian and $\vert j$ the 
covariant derivative on the three-sphere. With respect to
reparametrisations of the three-sphere, $h$, $B$ and
$E$ are scalars, $V_{i}$ and $F_{i}$ are divergenceless vectors
and $t_{ij}$ is a transverse traceless symmetric tensor.

One may expand the spatial part of each of these spin-$r$ fields in terms of a 
complete set of harmonics, labelled by the eigenvalues 
of the Laplacian on $S^3$, $\lambda_{p} =p^2 +(r+1)$
where $p=in$ and $n$ is an integer. 
In general the decomposition of a metric perturbation $h_{ij}$ 
into a scalar, vector 
and tensor part is unique. Hence one can write $E$, $F_{i}$ and 
$t_{ij}$ back in terms of $h_{ij}$ \cite{Kodama}. 
For scalar $p^2_{s}=-4$ and vector $p^2_{v} =-4$ harmonics however, 
the decomposition is not unique and there
appears 
a degeneracy between 
scalar- or vector-type perturbations  
and $p^2_{t}=0$ and $p^2_{t}=-1$ 
tensor modes respectively. Treatment of the former is complicated 
by the involvement of the 
scalar field \cite{Gratton}, but the latter mode is
unambiguously
pure gauge. 
We will return  to this point in section 4.

The Euclidean action is
\begin{equation}\label{action}
S = \frac{1}{2\kappa} \int d^4 x \sqrt{g}\left( -R + \frac{1}{2}
\nabla_{\mu} \phi \nabla^{\mu} \phi + V(\phi)\right) -
\frac{1}{\kappa}\int d^3 x \sqrt{g} K,
\end{equation}
where the surface term is needed to remove 
second  derivatives.
Substituting the decomposition (\ref{dec}) into the action (\ref{action}),
we keep 
all terms to second order. 
The scalar, vector and 
tensor quantities decouple. The 
scalar perturbations are studied in 
 ref. \cite{Gratton}. The 
vector perturbations are uninteresting to first order in $\hbar$ since they are forced to
be zero by the Einstein constraints. The tensor perturbations 
give the following second order positive Euclidean action
\begin{equation}\label{action2}
S_{2} = 
\frac{1}{8\kappa}
\int d^4 x \sqrt{\gamma} b^2 \left(
t'^{ij}t'_{ij} + t^{ij\vert k}t_{ij\vert k} + 2 t^{ij}t_{ij} \right),
\end{equation}
where  prime denotes differentiation with respect to the conformal coordinate 
$X$. If one performs the rescaling
$\tilde t_{ij} = b(X)t_{ij}$, and integrates by parts 
one obtains
\begin{equation}\label{act}
S_{2} = 
\frac{1}{8\kappa}
\int d^4 x \sqrt{\gamma}
\tilde t_{ij} \left(
 \hat K+3 -\Delta_{3} \right)\tilde t^{ij} -{1\over 8\kappa} \left[\int d^3x \sqrt{\gamma}  \tilde 
t_{ij}\tilde t^{ij} {b'\over b}(X)\right]
\end{equation}
where the Schr\"{o}dinger operator 
\begin{equation}\label{pot}
\hat K = -\frac{d^2}{dX^2}  + \frac{b''}{b} -1 \equiv  -\frac{d^2}{dX^2}  +U(X)
.\end{equation}

The form of the potential $U(X)$ is shown in Figure 1 for de Sitter space,
as well as examples of a Coleman-De Luccia instanton and a Hawking-Turok instanton.
The operator $\hat K$ has in all three cases
a positive continuum starting at eigenvalue
$p^2 = 0$, as well as a single bound state $\tilde t_{ij} = b(X)q_{ij}(\Omega)$
at $p=i$.

For singular instantons the surface terms in (\ref{act}) play a crucial 
role. The potential $U(X) \rightarrow -{1\over 4 X^2}$ as $X\rightarrow 0$. 
The eigenmodes of $ \hat K$ 
behave
as $X^{1/2}$ or $X^{1/2}\mathrm{ln} X$ 
near the singularity. The latter modes contribute positive infinity
to the surface terms in (\ref{act}), and are therefore suppressed 
in the path integral. 
Hence we see that as in the scalar case \cite{Gratton}, the path integral 
unambiguously specifies the allowed fluctuation modes as those which vanish at
the singularity.

\hfill\break 
\begin{figure}
\centerline{\psfig{file=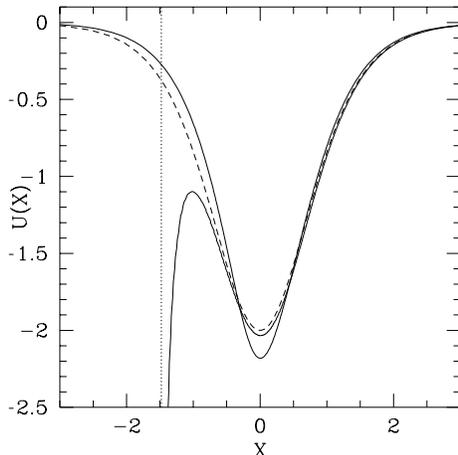,width=2.5in}}
\caption{Potential $U(X)$ occurring in the 
Schr\"{o}dinger operator governing tensor perturbations 
about the various instanton solutions discussed in the text.
The dashed line shows the potential
for an $S^4$ instanton corresponding to perfect de Sitter space, 
where $U(X)= -2/\cosh^{2}(X)$.
The upper solid line shows the potential for a
Coleman-De Luccia instanton, where $-\infty <X <\infty$, 
and the
lower solid line that for 
a 
Hawking-Turok instanton, with singularity indicated by the vertical
dotted line.  
The potentials have been shifted in $X$ so their minima 
coincide.
All three are very similar to the right
of the minimum. To the left, the Hawking-Turok potential diverges 
as one approaches the singularity. The potential is reflectionless
in the $S^4$ case,  weakly reflecting in the Coleman-De Luccia
case and totally reflecting in the Hawking-Turok case.
} 
\end{figure}

\section{The Euclidean Green Function}

To evaluate the path integral (\ref{wave}), we first 
look for the Green function 
$G_{E}^{\ iji'j'}$
of the operator in (\ref{act}). The Euclidean fluctuation 
correlator (\ref{wave})
will then be given by $b^{-1}(X) b^{-1}(X') G_{E}^{\ iji'j'}$.
The Euclidean Green function satisfies
\begin{equation}\label{green}
\frac{1}{4\kappa} \left(
\hat K +3 -\Delta_{3} \right)G^{\ ij}_{E\ i'j'}
(X,X',\Omega , \Omega ')
=\delta (X-X') \gamma^{-{1\over 2}} \delta^{ij}_{\ \ i'j'}(\Omega - \Omega ').
\end{equation}
If we think of the scalar product as defined by integration over $S^3$
and summation over tensor indices, then 
the 
right hand side is the normalised projection operator onto
transverse traceless tensors on $S^3$.
Since the eigenmodes of the Laplacian 
form a
complete basis, we can write the last term as
\begin{equation}
\delta^{ij}_{\ \ i'j'}(\Omega - \Omega ') = \sum_{k} \sum_{{\cal P}=e,o}
\sum_{l = 2}^{k}\sum_{m  = -l}^{l}q^{(k)ij}_{{\cal P}lm}(\Omega )
q^{(k),{\cal P}lm}_{i'j'}(\Omega ')^{*}
\end{equation}
where on $S^3$ we have
\begin{equation}
 \Delta_{3} q^{(k)ij}_{{\cal P}lm} = \lambda_{k}q^{(k)ij}_{{\cal P}lm}
\end{equation}
with $\lambda_{k} = -k(k+2) + 2$. Here ${\cal P}=\{e,o\}$ labels the parity,
the angular momentum on $S^3$ takes the
values $k = 2,3,4,..$ and
$2\leq l\leq k $ and $-l \leq m \leq l$ are the usual quantum numbers on the two-sphere.
Note that $(l\geq 2)$ because a spin-two field has no
monopole or dipole components. The eigenmodes are normalized by the condition
\begin{equation}\label{norm}
\int \sqrt{\gamma} d^{3} x q^{(k)ij}_{{\cal P}lm}
q_{{\cal P}'l'm'ij}^{(k')*} = \delta^{kk'}\delta_{{\cal P}{\cal P}'}\delta_{ll'}\delta_{mm'}
\end{equation}

The set of eigenmodes
form a representation of the symmetry group $SO(4)$ of the
manifold. It follows in particular that 
the sum over ${\cal P},l$ and $m$ defines a maximally symmetric 
bitensor \cite{Jacob}
\begin{equation}\label{bit}
W^{\ ij}_{(k)\ i'j'}(\mu) =
\sum_{{\cal P}lm} q^{(k)ij}_{{\cal P}lm}(\Omega ) q^{(k){\cal P}lm}_{i'j'}(\Omega ')^{*}
\end{equation}
that depends only on the geodesic distance 
$\mu (\Omega , \Omega')$ between the two spacetime points.
The Green function $G_{E\ i'j'}^{\ ij}$ can only be a function of 
$\mu (\Omega , \Omega')$ if it is to be invariant under isometries of the
three-sphere. 
Note that the indices $i,j$ lie in the tangent space over the point $\Omega$
while the indices $i',j'$ lie in the tangent space over the point $\Omega'$.
The general form of the bitensor $W^{\ ij}_{(k)\ i'j'}$ 
appearing in tensor fluctuation correlators
has been obtained by Allen \cite{Allen} and is given
in Appendix A below. 
Here we note already that in terms of the label $p=i(k+1)=in$,
the bitensor on $S^3$ has precisely 
the same formal expression as the corresponding 
object on $H^3$. Since we would like to analytically continue our 
result for the Euclidean two-point
correlator into the
open universe, we will use the label $p=in$ 
from now on. 
We now return to equation (\ref{green}) for the the Euclidean Green function.

By substituting the following ansatz for the Green function,
\begin{equation}
G^{\ ij}_{E\ i'j'}(\mu,X,X') = 
4\kappa \sum_{p=in} G_{p}(X,X') W^{\ ij}_{(p)\ i'j'}(\mu ),
\end{equation}
into (\ref{green}) and noting that in terms of $p=in$, we have
$\lambda_k= p^2+3$, we obtain an
equation for the 
model-dependent part of the Green function,
\begin{equation}\label{xgreen}
\left(\hat K -p^2\right) G_{p}(X,X') = \delta (X-X')
\end{equation}

Let us first discuss the case of singular instantons.
The solution to equation (\ref{xgreen}) is 
\ba
 G_{p}(X,X') &=&
{1\over \Delta_p} \left[\Psi_{p}^{+}(X) \Psi_{p}^{-}(X')\Theta(X-X')+
\Psi_{p}^{-}(X) \Psi_{p}^{+}(X')\Theta(X'-X)\right],
\ea
where  $\Psi_{p}^{-}(X)$ is the solution to the Schrodinger 
equation that goes as $X^{1/2}$ as $X \rightarrow 0$ and 
$\Psi_{p}^{+}(X)$ is the solution going as $e^{ipX}=e^{-nX}$ 
as $X$ tends to infinity.  The factor $\Delta_p$ is 
the Wronskian $ {\Psi_{p}^{-}}'\Psi_{p}^{+}-{\Psi_{p}^{+}}'\Psi_{p}^{-}$ 
of the two solutions.

We shall ultimately be interested in re-expressing this solution 
as an integral over real values of 
$p$ in order to continue it to the open universe.
To do so we must extend the 
solutions $\Psi_{p}^{\pm}$ 
defined above at $p=in$ into the complex  $p$-plane.
$\Psi_{p}^{-}(X)$ becomes $\Psi_{p}(X)$, 
defined for all complex $p$ to be 
the solution which tends to $X^{1\over 2}$
as
$X \rightarrow 0$. Being a solution of a regular differential
equation this is analytic for 
finite $p$ in the complex  $p$-plane. On the other hand,
$ \Psi_{p}^{+}(X)$ is the analytic continuation of
$g_{p}(X)$, defined on the real $p$ axis to be the solution 
tending to  $e^{ipX}$ 
as $X \rightarrow \infty$. This is the Jost function,
and is analytic in the upper half $p$-plane \cite{newton}.
The two solutions may be expressed in terms of
each other as 
\begin{equation}\label{decomp}
\Psi_{p}(X) = a_{p}^{\ }g_{p}(X) + a_{-p} g_{-p}(X),
\end{equation}
and their 
Wronskian $\Delta_p = \Psi_{p}'g_p - g_p'\Psi_p=-2ipa_{-p}$,
independent of $X$.
This too is
analytic in the upper half $p$-plane. Zeros of $a_{-p}$ in the
upper half $p$-plane correspond to normalisable bound states. 
They can only occur on the imaginary $p$-axis, and in the case of
interest here the only zero  in the upper half $p$-plane
is at $p=i$. This zero corresponds to the 
bound state mentioned above.
For $X > X'$  we have the Green function as a discrete sum
\begin{eqnarray}
G_{E}^{\ iji'j'}(\mu,X,X') & = & 4\kappa \sum_{p=3i}^{+i\infty}
\frac{i}{2pa_{-p}^{\ }}
\Psi_{p}^{+}(X)\Psi_{p}^{-}(X')W^{\ iji'j'}_{(p)}(\mu)
\label{disc}
\end{eqnarray}

For regular Coleman-De Luccia instantons a similar procedure may be followed.
Here $X$ ranges from $-\infty$ to $+\infty$ and we define 
the two linearly independent 
mode
functions  $g_p^{\mathrm{left}}\(X\)$, which  tends to $e^{-ipX}$ as
$X\rar-\infty$, and $g_p^{\mathrm{right}}\(X\)$, which tends to
$e^{ipX}$ as $X\rar \infty$.
These can be shown to be orthogonal and
analytic in the upper half $p$-plane.
As $X \rightarrow +\infty$, we have
$g_p^{\mathrm{left}}\(X\) \rightarrow c_{p} e^{ipX} + d_{p}e^{-ipX}$.
Hence, the Wronskian 
$\Delta_p ={g_p^{\mathrm{left}}}'g_p^{\mathrm{right}}-{g_p^{\mathrm{right}}}'
g_p^{\mathrm{left}}= -2ipd_{p}$ and the Green function
$G_{E}^{iji'j'}(\mu,X,X')$ may be expressed in a form analogous to
that for 
singular instantons.


Before proceeding to the analytic continuation, let us demonstrate that
our Euclidean  Green functions are regular at the regular pole.
This is a nontrivial check because the coordinates $\sigma$ and $X$ are singular
there, and the rescaling becomes divergent too, $b(X)\sim \sigma^{-1} \sim e^{+X}$. 
In the large $X, X'$ limit, (\ref{disc}) becomes
\begin{equation}\label{sum}
G_{E}^{\ iji'j'}(\mu,X,X') = 
2 \kappa \sum_{n=3}^{\infty}\frac{1}{n}\left(
 e^{-n(X - X')} + \frac{a_{in}^{\ }}{a_{-in}}e^{-n(X+X')}\right)
W^{\ iji'j'}_{(in)}(\mu )
\end{equation}
For $n \geq 3$ the Gaussian hypergeometric functions  $F(3+n,3-n,7/2,z)$ that
constitute the bitensor $W^{\ iji'j'}_{(n)}$ have a series
expansion that terminates, and they 
essentially reduce
to Gegenbauer's polynomials $C^{(3)}_{n-3}(1 - 2z)$.
Using then the identity \cite{Erdelyi}
\begin{equation}
\sum_{l=0}^{\infty} C^{\nu}_{l}(x)q^{l} = \left( 1 -2xq + q^2 \right)^{-\nu}
\end{equation}
with $q = e^{-(X \pm X')}$, one easily sees that the sum
(\ref{sum}) indeed converges. 

We have the Euclidean Green function defined as an infinite sum \ref{disc}).
We wish to represent it as an integral over $p$. To do so
we must extend the summand into the upper half $p$-plane. We have already 
defined the
wavefunctions for all complex $p$ but we need to extend the 
bitensor as well. When the Green function is expressed as a discrete sum,
it involves the bitensor $W^{\ iji'j'}_{(p)}(\mu )$ evaluated at $p=in$ with
$n$ integral. At these values of $p$, the bitensor is regular at both
coincident and opposite points on $S^3$, that is at $\mu=0$ and $\mu=\pi$.
However, if we extend $p$ 
into the complex plane we lose regularity at 
$\mu=0$. This is clear from (\ref{green}). For if we distort
the $p$ integral to run along the real axis, and use the completeness relation
for the eigenfunctions $\psi_p(X)$, it follows that  $W^{\ iji'j'}_{(p)}(\mu)$ 
obeys a differential equation with a delta function source at $\mu=0$
(see the discussion of the scalar case in \cite{Gratton}). 
Similarly, when 
we extend $W^{\ iji'j'}_{(in)}(\mu )$ into the complex $p$-plane, 
we must maintain regularity at $\mu=\pi$, since there is no delta
function source there. 

The condition of regularity at $\pi$ imposed by the differential
equation for the Green function is sufficient
to uniquely specify the analytic continuation of $W^{\ iji'j'}_{(in)}(\mu )$
into the complex $p$-plane. To see this, we note from Appendix $A$ that 
The bitensor involves coefficent functions $\alpha$ and $\beta$ which
are hypergeometric functions of the variable $z=\cos^2(\mu/2)$.
For coincident points, $z=1$ but for antipodal points $z=0$. There
are two independent solutions of the hypergeometric equation,
namely $\alpha(z)$ and $\alpha(1-z)$. They
are related by the transformation formula
(eq.[15.3.6] in \cite{Abram})
\ba
&&_2F_{1}(3+ip,3-ip,{7\over 2} ,z)  = 
(-\cosh p\pi) _2F_{1}(3+ip,3-ip,{7\over 2},1-z)\nonumber\cr
&& +
\frac{\Gamma ({7\over 2})\Gamma ({5\over 2})}{\Gamma(3+ip)\Gamma(3-ip)}(1-z)^{-{5\over 2}}
\ _2F_{1}({1\over 2} -ip,{1\over 2} +ip,-{3\over 2} ,1-z)
\label{transf}
\ea
Notice that for the eigenvalues of the Laplacian on $S^3$, i.e.
$p=in\ (n\geq 3)$,
the second term on the right-hand side vanishes. In this case the two choices
are simply related by $(-1)^{n+1}$ and they are both regular
for all $\mu$. Since $F(1-z) \rightarrow 1$
for coincident points, we must take this solution in (\ref{disc}).
But when we express the discrete sum (\ref{disc}) as a contour
integral, to maintain regularity of the integrand at $\mu=\pi$ we need to
first replace $F(1-z)$ by a term $F(z)(-1)^{n+1}$, and then continue the
latter term 
to $- (\cosh p\pi)^{-1}
\ _2F_{1}(3+ip,3-ip,{7\over 2} ,z)$.

Now we write the sum in (\ref{disc}) as an integral 
along a contour ${\cal C}_1$ encircling
the points $p=3i, 4i, ... Ni$ on the imaginary $p$-axis,
 where $N$ tends to infinity. 
Using the analytic properties
of the terms in the discrete sum extended into the
complex $p$-plane we have for $X>X'$, 
\begin{eqnarray}
G_{E}^{\ iji'j'}(\mu,X,X') & = & \kappa\int_{{\cal C}_1} {dp \over p
\sinh p \pi} 
{g_p(X) \psi_p(X') \over a_{-p}} 
W^{\ iji'j'}_{(p)}(\mu).
\label{discint}
\end{eqnarray}
where $W^{\ iji'j'}_{(p)}(\mu)$ is defined in the Appendix,
equations (A6), (A10) and (A11),
using the forms regular at $\mu=\pi$ i.e.
$\alpha(z)$ and $\beta(z)$. From the explicit forms 
given, it is clear that $W^{\ iji'j'}_{(p)}(\mu)$ is
analytic in complex $p$-plane in the required region.
To see that (\ref{discint}) is equivalent to the sum (\ref{disc}) 
introduce $1=\cosh p \pi /\cosh p \pi$ into the integral.
Then note that 
$\coth p\pi$ has 
residue $\pi^{-1}$ at every integer
multiple of $i$. Finally, use (\ref{transf}) at $p=ni$ to
rewrite $W^{\ iji'j'}_{(p)}(\mu)$ in the form regular 
at $\mu=0$. The factor of $\cosh p \pi$ from (\ref{transf})
cancels that in the integrand. Note also the minus sign that appears 
in (\ref{transf}) cancels that introduced by the change in
sign of the normalisation factor $Q_p = -(p^2+4)/(30 \pi^2)$,
which is positive if $p=in,$ $n\geq 2$, but negative for real $p$.
The cancellation of these signs ensures that the
Lorentzian correlator has the correct positivity properties. 

We now distort the contour for the $p$ integral to run 
along the real $p$ axis (Figure 2). At large imaginary $p$
the integrand decays exponentially 
and the contribution vanishes in the limit of large $N$. 
However as we deform the contour towards the real axis 
we encounter two poles in the $\coth p\pi$ factor, the 
latter at $p=i$ becoming a
double pole due to the simple zero of $a_{-p}$. 
For the $p=2i$ pole, we note that it follows directly from the
the normalisation factor $Q_p$ 
that
$W_{(2i)}^{iji'j'}=0$. Indirectly, this is a consequence of 
the fact that spin-2 perturbations do not have a monopole or dipole component.
At $p=i$ we have a double pole. However, the bound state wavefunction is just 
proportional to $b(X)$ and the metric tensor perturbation $t_{ij}=
b^{-1}(X) \tilde t_{ij}$ is therefore independent of $X$. The latter coordinate
continues to conformal time in the open universe, and it follows that
the metric perturbation is time-independent and will not contribute to
the Sachs-Wolfe formula
(\ref{temp}). However to understand this mode more deeply, recall that
for
$p^2 = -1$ a degeneracy appears between $p^2=-1$ tensor-type perturbations and
$p_{v}^2=-4$ vector-type perturbations \cite{Tanaka}. To be more precise,
the tracelesss transverse tensors $q_{ij}^{(i)plm}$ may be constructed
from the vector harmonics $V_{i}^{(2i)plm}$ by
symmetrised covariant differentiation. One therefore has
$q_{ij}^{plm}(p^2=-1) = V_{(i \vert j)}^{plm}(p^2_{v}=-4)$.
This means that this discrete tensor mode is not invariant
under 
(vector) gauge transformations. It may be generated
by a purely spatial gauge transformation without disturbing the value of
the scalar field \cite{Tanaka}.
We may therefore use the remaining
gauge freedom in the decomposition (\ref{dec}) to set 
$W^{iji'j'}_{(i)}=0$.
We conclude that up to a term involving a pure 
gauge mode, we can deform the contour ${\cal C}_1$ into
the contour ${\cal C}$ shown in Figure 2. 
Since the integrand involves a factor $(p \sinh p\pi)^{-1}$
 which 
has a double pole at $p=0$, we leave the contour avoiding the origin 
on a small semicircle in the  upper half $p$-plane. We
shall see that for the Coleman-De Luccia and Hawking-Turok cases 
the complete integrand is actually regular at $p=0$, but
for perfect de Sitter space the double pole survives. 
The contribution to the Green function from the small semicircle 
acts to regulate the integral $\int_0^\infty dp/p^2$ coming 
from the real axis. Thus in our treatment, even in perfect de Sitter
space the Green function is finite, in contradiction to the
conclusion reached in treatments based on mode-by-mode matching. 

\begin{figure}
\centerline{\psfig{file=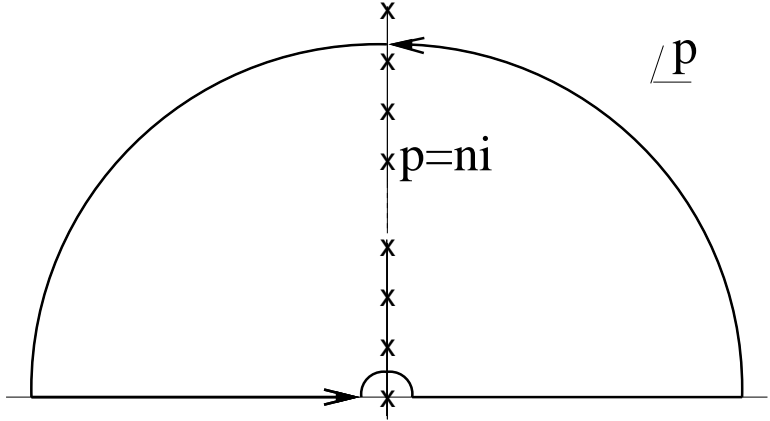,width=4in}}
\caption{Contour for the Correlator.}
\end{figure}

\section{Two-Point Tensor Correlator in an Open Universe}

The analytic continuation into the open universe is given by
setting 
$\Omega = -i \chi$ and $\sigma = it$ (see \cite{Gratton}) and letting
$a(t) \rightarrow b(\sigma ) \equiv -ia(i\sigma)$.
Here $\Omega$ is the polar angle on the three-sphere. For our
correlator, without loss of generality 
we may take one of the two points, say $\Omega'$ to be at the
north pole of the three-sphere. Then $\mu=\Omega$, and 
$\mu$ continues to $-i\chi$. We then obtain the correlator
in the open universe where one point has been chosen
as the origin of the radial coordinate $\chi$. 

The background line element of the Lorentzian region is
\ba
ds^2 = -dt^2 + a^2(t) \left( d\chi^2 + \sinh^2 \chi d\Omega_2^2 \right).
\ea
The conformal coordinate $X$ continues to conformal time $\tau$ 
as follows
\ba
X \equiv \int_{it}^{\sigma_{m}} \frac{d\sigma}{b(\sigma)} = - \tau - \frac{i
\pi}{2}
\label{contx}
\ea
where the conformal time $\tau$ is defined via
\ba
\ \tau \equiv \lim_{\epsilon \rightarrow 0} \left( \int_{\epsilon}^{\sigma_{m}}
\frac{d\sigma}{b(\sigma )} - \int_{\epsilon}^{t}\frac{dt'}{a(t')}
\right).
\label{taudef}
\ea
We now wish to make the substitutions $\mu = -i \chi$, where
$\chi$ is the comoving separation on $H^3$,
in the open universe, and $X=-i{\pi \over 2} -\tau$. 
The first continuation may be
done immediately.
We use the explicit formula for the bitensor regular at
$\mu=\pi$, given in the Appendix,
equations (A6), (A10) and (A11) to write the 
in following $p$-integral for the Euclidean Green function:
\begin{eqnarray}
\label{eucor}
G_{E}^{\ iji'j'}(\mu,X,X')
& = & 
\kappa         
\int_{C}\frac{dp}{p \sinh p\pi}
\left(
g_{p}(X)g_{-p}(X')+\frac{a_{p}}{a_{-p}} g_{p}(X) g_{p}(X') \right)
W_{\ iji'j'}^{(p)}(\chi)
\end{eqnarray}
where we have used the formula (\ref{decomp}) to re-express $\psi_p$ in terms
of the Jost functions $g_p(X)$. 
The obstacle to setting $X= - \tau - \frac{i
\pi}{2}$ is that
the integrand of (\ref{eucor}) contains a term $g_p(X) g_p(X') \sim e^{ip(X+X')}$.
If we simply make the substitution $X=-i{\pi \over 2} -\tau$ this would produce a term
going as $e^{p\pi}$. But the bitensor defined in (A10) and (A11)
involves terms which behave as 
$e^{+p (\pi +i\chi)}$, and the two factors of 
$e^{p\pi}$
would lead to  a meaningless divergent integral.
To circumvent the problem, we use the following identity. 
For $X-X'>0$, we have 
\ba
\int_{\cal C} {dp \over p} 
{g_p(X) \psi_p(X') \over a_{-p}} e^{i p\chi} F(p)
= 0
\label{intid}
\end{eqnarray}
where $F(p)$ are the $p$-dependent
coefficients occurring in the final (Lorentzian) form for the
bitensor given in (A12).
This identity 
follows from the analyticity
properties of the integrand explained above, and
the fact that, despite first appearances, the formulae
$A12$ are actually analytic at $p=i$. 
We now insert $1= \sinh p\pi/ \sinh p\pi$ under the integral, to
show that 
the integral (\ref{intid}) with  a factor $e^{p\pi}/\sinh p \pi$ 
inserted equals that with a factor $e^{-p\pi}/\sinh p \pi$ inserted. 
The resulting identity allows us to replace the dangerous terms
in the bitensor $e^{+p (\pi +i\chi)}$ by $e^{-p \pi +ip\chi}$.

We now perform the $X$ continuation. 
The analytic continuation of the Euclidean mode functions is
given by 
\ba
g_{\pm p}(X) \rightarrow e^{\pm p \pi \over 2} g^L_{\pm p}(\tau)
\labeq{jostcont}
\ea
where the Lorentzian Jost function $g_p^L(\tau)$ is the
solution to the Lorentzian perturbation equation
$\hat{K} g_p^L(\tau)
= p^2 g_p^L(\tau)$ obeying $g_p^L(\tau) \rar e^{-i p\tau}$ as
$\tau \rightarrow -\infty$. Equation (\ref{eq:jostcont}) follows
by matching
at large $X$. 
We finally obtain the Lorentzian tensor Feynman (time-ordered)
correlator, for $\tau'-\tau$, 
\begin{eqnarray}
\label{lorcor}
G_{L}^{\ iji'j'}(\chi,\tau,\tau')
& = & 
\kappa         
\int_{C}\frac{dp}{p \sinh p\pi}  \left( e^{-p\pi} 
g_{p}^{L}(\tau )g_{-p}^{L}(\tau ')+\frac{a_{p}}{a_{-p}}
g_{p}^{L}(\tau)g_{p}^{L}(\tau ')\right)
W_{\ iji'j'}^{L (p)}(\chi)
\end{eqnarray}
where the Lorentzian bitensor $W_{\ iji'j'}^{L (p)}(\chi)$ 
of relevance in the hyperbolic universe is defined
in the Appendix, equation (A12).
The factor $a_{p}/a_{-p}$ is  simply
a phase, since for real $p$ the Euclidean wavefunction is real so
$a_p^*=a_{-p}$.

Now we would like  to represent the result (\ref{lorcor}) 
as an integral over real $0<p<\infty$. The term $p^{-1} \coth p\pi$
in the integrand seems to produce a double pole at $p=0$. However,
for either the Coleman-de Luccia or Hawking-Turok instantons, 
the reflection term in (\ref{lorcor}) turns out to precisely
cancel the first term as $p \rightarrow 0$. This cancellation
seems to have first been discovered in refs. 
\cite{Sasaki}, \cite{Cohn}. The reason for the cancellation 
is that for any potential except a perfectly reflectionless one, 
at very low momenta (i.e. very long wavelengths) the wavefunction 
is completely reflected. This means that in the small 
$p$ limit both $a_{p}/a_{-p}$ and $c_p/d_p$ tend to minus one
\cite{garrigadisc}. 
By analyticity, we expect them to vanish as $p^2$, which
makes the integrand of (\ref{lorcor}) analytic as $p\rightarrow 0$.
It is however clear from the form of the potentials (Figure 1)
that the Coleman-De Luccia instantons are much closer to
the perfect $S^4$ non-reflecting solution. Therefore we 
may expect the regime  $c_p/d_p\rightarrow -1$ to set in at much
lower $p$ than in the Hawking-Turok case. This will lead to a 
larger contribution to the large angle microwave anisotropies. 
As mentioned above, a virtue of our treatment seems to
be that even the de Sitter result is finite. 

In the cases of interest therefore there is no singularity
at $p=0$, and we may take the contour to run along the 
real $p$-axis. Using the symmetry $p \rightarrow -p$, 
The right hand side of (\ref{lorcor}) 
becomes 
\ba
{\kappa\over 2} \int_{-\infty}^\infty & & {dp \over p}
W_{\ iji'j'}^{L(p)}(\chi)
\Bigg( {\rm coth}  p\pi \[
g_p^L(\tau) g_{-p}^L(\tau') +g_{-p}^L(\tau) g_{p}^L(\tau')\] \cr
 -&& \[
g_p^L(\tau) g_{-p}^L(\tau') -g_{-p}^L(\tau) g_{p}^L(\tau')\]
+  {1\over {\sinh}  p\pi} \[
{a_p \over a_{-p}} g_p^L(\tau) g_{p}^L(\tau')+
{a_{-p} \over a_{p}} g_{-p}^L(\tau) g_{-p}^L(\tau')\]
\Bigg).
\labeq{rewriii}
\ea
For real $p$,  $g_{-p}(\tau)^L$ is the complex conjugate of $g_{p}^L(\tau)$
and $a_{-p}$ of $a_p$. So the second term is imaginary but the 
first and third terms are real. In fact it is straightforward to
see that if we apply the Lorentzian version of the perturbation operator
$\hat K$ to (\ref{eq:rewriii}) with an appropriate heaviside function of
$\tau-\tau'$, the imaginary term will produce the Wronskian of $g_{-p}(\tau)^L$
and $g_{p}^L(\tau)$, which is proportional to $p$,
  times $\delta(\tau-\tau')$. Then the integral over $p$ produces a spatial
delta function. From this one sees that our
Feynman correlator obeys the correct
second order partial differential equation, with a delta function source.
The delta function goes from being real in the Euclidean region
to imaginary in the Lorentzian region because of the factor $\sqrt{\gamma}$ 
in (\ref{green}).

For cosmological applications, we are usually interested in the 
expectation of some quantity squared, like the microwave background 
multipole moments.
For this purpose, all that matters is the symmetrised correlator
$\langle \{ t_{ij}(x),t_{i'j'}(x')\}\rangle$ which is just the real part
of the Feynman correlator. It also represents the `classical'
piece, which in the situations of interest, where occupation numbers
of modes are large, is much larger than the quantum piece.

For the tensor correlator we 
also need to restore the factor  $a^{-1} (\tau)$ to $t_{ij}$.
It is convenient to define the 
eigenmodes $\Phi_{p}^{L}(\tau)= g_p^L(\tau)/a(\tau)$.
The symmetrised correlator is then given by
\begin{eqnarray}\label{lorcorf}
\langle \{ t_{ij}(x),t_{i'j'}(x')\}\rangle & = &
2\kappa \Re
\int_0^\infty \frac{dp}{p}\left(\coth p\pi
\Phi_{p}^{L}(\tau )\Phi_{-p}^{L}(\tau ')+\frac{a_{p}^{\ }}{a_{-p}^{\ }}
\frac{\Phi_{p}^{L}(\tau)\Phi_{p}^{L}(\tau ')}{\sinh p\pi }\right)
W_{\ iji'j'}^{L (p)}(\chi).
\end{eqnarray}
where $W_{\ iji'j'}^{L(p)}(\chi)$ is defined in the Appendix, equations
(A3) and (A12). 

In this integral the bitensor $W^{\ iji'j'}_{(p)}(\mu)$
equals the sum of the rank-two tensor eigenmodes with
eigenvalue $\lambda_{p} =-(p^2+3)$ of the Laplacian on $H^3$.
At large $p$, its coefficient functions $w_{j}^{(p)}$
(see Appendix A) behave like $p\sin p\mu$. Hence
the above integral 
converges at large $p$ for both timelike and spacelike 
separations. Equation (\ref{lorcorf}) is our final result
for the tensor spectrum from singular instantons. 
As in the scalar calculation \cite{Gratton}, 
and as mentioned above, for Coleman-De Luccia instantons the 
phase $a_p/a_{-p}$ gets replaced by $c_p/d_p$, which is the reflection
amplitude for waves incident from $X=+\infty$ in
the Euclidean region.

Before moving on to the observational consequences of  (\ref{lorcorf}) 
we would like to make one more technical comment.
We mentioned already that a degeneracy appears
between $p^2 =0$ tensor modes and $p_{s}^2 =-4$ scalar
perturbations. These discrete modes were initially 
interpreted 
as bubble wall fluctuations \cite{Garr,Bellido}. However, 
in our approach they do not  contribute in the scalar calculation 
(for $l\geq 2$) because the corresponding spherical harmonics
are singular and overcomplete on the Euclidean three-sphere. 
More recently the wall fluctuations were argued to
have re-appeared as a 
long-wavelength continuum contribution on top of 
the usual continuous spectrum of even parity
gravitational wave modes \cite{Sasaki}. In this way, 
the bubble wall fluctuations were found to
regularize the tensor spectrum, thought to be infrared divergent
in pure de Sitter space \cite{Sasaki}.
Our result for the correlator for a Coleman-De Luccia model is indeed infrared finite
and the cancellation caused by total reflection of low
momentum modes allowed us to represent the result as an integral 
starting at $p=0$. However we do not agree that the presence of
the bubble was needed to 
regularize the spectrum. In our method, even in perfect de Sitter
space we obtain a finite result, because the contribution of
the small semicircle on the contour ${\cal C}$ shown in Figure 2
regularises the final answer. So in our approach
the tensor spectrum in perfect de Sitter space appears to be
infrared finite, contrary to the findings of earlier works.
This issue is academic for present purposes,  but clearly
 deserves further study.

\section{Implications for the CMB-anisotropy}

Gravitational waves provide an extra source of
time-dependence in the background in which the cosmic microwave
background photons propagate. 
The contribution of gravitational waves to the
CMB anisotropy is given by the integral in the Sachs-Wolfe 
formula \cite{Sachs},
\begin{eqnarray}\label{temp}
\frac{\delta T_{SW}^{\ }}{T}(\theta,\phi)& = &
-\frac{1}{2}\int_{\tau_{e}}^{\tau_{0}}d\tau
t_{\chi \chi,\tau}^{\ }(\tau,\chi,\theta,\phi)|_{\chi = \tau_{0} -\tau}
\end{eqnarray}
where $\tau_{0}$ and $\tau_{e}$ are respectively the observing and
last scattering time for the photons and $\chi$ is the 
comoving radial coordinate.
The anisotropy is characterised by the two-point angular correlation function
$C(\gamma)$, where $\gamma$ is the angle
between two points on the celestial sphere.
It is customary to expand the correlation function in terms of
Legendre Polynomials as
\begin{equation}\label{ang}
C(\gamma) =\left\langle \frac{\delta T}{T}(0)\frac{\delta T}{T}(\gamma)
\right\rangle =
\sum_{l=2}^{\infty} \frac{2l+1}{4\pi} C_{l} P_{l}(\cos \gamma)
.\end{equation}
where in standard notation $C_{l} = \langle \vert a_{lm} \vert^2 \rangle$.
Hence, inserting the Sachs-Wolfe integral into (\ref{ang}) and substituting
(\ref{lorcorf}) for the two-point fluctuation correlator yields
\begin{equation}
C(\gamma)  = 
\frac{1}{4}\int_{\tau_{e}}^{\tau_{0}}d\tau
\int_{\tau_{e}}^{\tau_{0}}d\tau' \frac{\partial}{\partial \tau}
\frac{\partial}{\partial \tau'}\langle
t_{\chi\chi}(\tau,0)t_{\chi'\chi'}(\tau',\gamma)\rangle
.\end{equation}
In order to obtain 
$C_{l}$
we write the bitensor back in terms
of its defining tensor eigenmodes on $H^3$. Since $q_{\chi\chi}^{(p)olm}=0$,
only the
even parity modes contribute to the CMB-anisotropy. 
The normalised eigenfunctions $q_{\chi\chi}^{(p)elm}(\chi,\theta,\phi)$
can be written as
$Q^{pl}_{\chi\chi}(\chi)Y_{lm}(\theta, \phi)$, where \cite{Tomita},
\begin{equation}
Q^{pl}_{\chi\chi}(\chi) = \frac{N_{l}(p)}{p^2(p^2+1)} 
(\sinh \chi)^{l-2} \left(\frac{-1}{\sinh \chi}\frac{d}{d\chi}\right)^{l+1}
(\cos p\chi)
\end{equation}
and
\begin{equation}
N_{l}(p) =\left[\frac{(l-1)l(l+1)(l+2)}{\pi \prod_{j=2}^{l}(j^2+p^2)}
\right]^{1/2}.
\end{equation}
Hence we obtain for the power spectrum of multipole moments
\begin{eqnarray}\label{multi}
C_{l} = 
\kappa \Re \int_{0}^{+\infty}\frac{dp}{2p}
\int_{\tau_{e}}^{\tau_{0}}d\tau
\int_{\tau_{e}}^{\tau_{0}}d\tau'
&& \left( \coth p\pi
\left[\dot\Phi_{p}^{L}(\tau )\dot\Phi_{-p}^{L}(\tau')\right] \right. \nonumber\\
&& \quad \left.+  \frac{1}{\sinh p\pi}
\left[\frac{a_{p}^{\ }}{a_{-p}^{\ }}\dot\Phi_{p}^{L}(\tau )
\dot\Phi_{p}(\tau')\right]\right)Q^{pl}_{\chi\chi}
Q^{pl}_{\chi'\chi'} 
\end{eqnarray}

The contribution to the multipole moments 
due to the second, reflection term 
falls exponentially with increasing wavenumber.
However in contrast with the scalar fluctuations the long-wavelength
tensor perturbations do give a substantial contribution
to the CMB anisotropies.
Hence the dependence of the tensor spectrum on the boundary conditions 
for the perturbations defined by the instanton background -
Dirichlet for Hawking-Turok, free boundary conditions for
Coleman-De Luccia, may
provide a way to observationally  distinguish different versions of
open inflation. From the discussion above, 
we expect a larger
contribution at low $p$ for regular instantons.
We shall perform the numerical computation of the needed reflection 
coefficients in future work \cite{gratthertog}.

In addition, for a complete calculation of
the $C_l$ one must evolve the Lorentzian mode functions 
$\Phi_p^L(\tau)$
forward from the beginning  $\tau=-\infty$ of inflation inside
the open universe up to the present time $\tau=\tau_0$. 
In the 
inflationary phase of the open 
universe the mode functions closely follow
perfect de Sitter evolution in which they tend to 
a constant after the physical wavelength has been stretched outside
the Hubble radius. The amplitude and phase of this 
constant defines initial conditions for the 
radiation and matter dominated eras in which the modes of interest
re-enter the Hubble radius. The radiation and matter evolution 
is straightforward to study numerically, and from this one can 
compute the 
Sachs-Wolfe integral (\ref{multi}) 
and the the multipole moments $C_l$.

\section{Conclusion}

We have computed the spectrum of tensor perturbations predicted in
open inflation, according to Euclidean no boundary initial 
conditions. The Euclidean path integral 
unambiguously specifies the tensor correlators with no additional
assumptions. We feel that the present  work places earlier results on
a substantially firmer footing. 
Our final result for the correlator, (\ref{lorcorf}),
and the cosmic microwave multipole moments (\ref{multi}) is given in
terms of scattering amplitudes in the Euclidean region and 
mode functions in the Lorentzian region. Both are straightforward to
compute numerically, and we shall do so in future work \cite{gratthertog}.

\bigskip
\centerline{\bf Acknowledgements}

We wish to thank Martin  Bucher and Steven Gratton for discussions, and
Stephen Hawking for advice and encouragement. One of us (NT) 
thanks J. Garriga for valuable comments. 

This work was supported by a PPARC (UK) rolling grant and an EPSRC studentship.
\hfill\eject

\appendix            

\section{Maximally Symmetric Bitensors}

A maximally symmetric bitensor $T$ is one for which $\sigma^{*}T=0$
for any isometry $\sigma$ of the maximally symmetric manifold.
Any maximally symmetric bitensor may be expanded in terms of a complete set of
'fundamental' maximally symmetric bitensors with the correct index symmetries.
For instance
\begin{eqnarray}\label{maxi}
T_{iji'j'} &  = & 
t_1(\mu) g_{ij}^{\ }g_{i'j'}^{\ }+
t_2(\mu)\left[n_{i}^{\ }g_{ji'}^{\ }n_{j'}^{\ }+
n_{j}^{\ }g_{ii'}^{\ }n_{j'}^{\ }+ n_{i}^{\ }g_{jj'}^{\ }n_{i'}^{\ }+
n_{j}^{\ }g_{ij'}^{\ }n_{i'}^{\ }\right]\nonumber\\
& & +t_3(\mu)\left[ g_{ii'}^{\ }g_{jj'}^{\ }+g_{ji'}^{\ }g_{ij'}^{\ }
\right]+ t_4(\mu)n_{i}^{\ }n_{j}^{\ }n_{i'}^{\ }n_{j'}^{\ }\nonumber\\
& & +t_5(\mu)\left[g_{ij}^{\ }n_{i'}^{\ }n_{j'}^{\ }+n_{i}^{\ }n_{j}^{\ }
g_{i'j'}^{\ }\right]
\end{eqnarray}
where the coefficient functions $t_{j}(\mu)$ depend only on the 
distance $\mu(\Omega,\Omega')$ along the shortest geodesic
from $\Omega$ to $\Omega'$.
$n_{i'}^{\ }(\Omega,
\Omega ')$ and $n_{i}^{\ }(\Omega, \Omega ')$ are
unit tangent vectors to the geodesics joining $\Omega$ and $\Omega'$ and
$g_{ij'}(\Omega, \Omega ')$ is the parallel propagator along the 
geodesic; $V^{i}g_{i}^{j'}$ is the vector at $\Omega'$ obtained by
parallel transport of $V^{i}$ along the geodesic from $\Omega$ to $\Omega'$
\cite{Jacob}.

The bitensor 
\begin{equation}\label{ana}
W^{\ ij}_{(p)\ i'j'}(\mu) =
\sum_{{\cal P}lm} q^{(p)ij}_{{\cal P}lm}(\Omega ) 
q^{(p){\cal P}lm}_{i'j'}(\Omega ')^{*}
\end{equation}
appearing in our Green function (\ref{bit}) has some additional 
properties arising from its construction in terms of the transverse and 
traceless tensor harmonics $q_{ij}^{(p){\cal P}lm}$ on $S^3$ (or $H^3$).
The tracelessness of $W^{(p)}_{iji'j'}$
allows one to eliminate two of the coefficient functions in
(\ref{maxi}).
It
may then be written as
\begin{eqnarray}\label{bitensor}
W^{(p)}_{iji'j'}(\mu) & = & 
w^{(p)}_1\left[ g_{ij}^{\ } -3n{i}^{\ }n_{j}^{\ }\right]
\left[g_{i'j'}^{\ } -n_{i'}^{\ }n_{j'}^{\ }\right]\ \ \ \ \ \ \ \ \ \ 
\cr
& & + w_2^{(p)}\left[n_{i}^{\ }g_{ji'}^{\ }n_{j'}^{\ }+
n_{j}^{\ }g_{ii'}^{\ }n_{j'}^{\ }+ n_{i}^{\ }g_{jj'}^{\ }n_{i'}^{\ }+
n_{j}^{\ }g_{ij'}^{\ }n_{i'}^{\ } +4n_{i}^{\ }n_{j}^{\ }n_{i'}^{\ }n_{j'}^{\ }
\right]\cr
& & 
+w_3^{(p)}\left[ g_{ii'}^{\ }g_{jj'}^{\ }+g_{ji'}^{\ }g_{ij'}^{\ }
 -2n_{i}^{\ }g_{i'j'}^{\ }n_{j}^{\ } -2n_{i'}^{\ }g_{ij}^{\ }n_{j'}^{\ }
+6n_{i}^{\ }n_{j}^{\ }n_{i'}^{\ }n_{j'}^{\ }\right]
\end{eqnarray}
The requirement that the bitensor be transverse
$\nabla^{i}W_{iji'j'}^{(p)}=0$ and the
eigenvalue condition $(\Delta_{3} - \lambda_{p})W^{\ iji'j'}_{(p)}=0$
impose additional constraints on the remaining coefficient functions
$w_{j}^{(p)}(\mu)$. To solve these constraint equations it is convenient to
introduce the new variables \cite{Allen}
\begin{equation}\label{bet}
\left\{
\begin{array}{lll}
\alpha(\mu) & = &  w_1^{(p)}(\mu) + 
w_2^{(p)}(\mu)\\
\beta(\mu)&  = & \frac{7}{(p^2 +9)\sin \mu}\frac{d\alpha(\mu)}{d\mu}
\end{array}
\right.
\end{equation}
where $\mu$ is the geodesic
distance on $S^3$.
In terms of a new argument $z=\cos^2(\mu/2)$ the transversality and eigenvalue conditions imply 
for $\alpha(z)$
\begin{equation}\label{hyper}
z(1-z)\frac{d^2\alpha(z)}{d^2z} + \left[ \frac{7}{2} -7z\right]
\frac{d\alpha(z)}{dz}=(p^2 +9)\alpha(z)
\end{equation}
and then for the coefficient functions
\ba
\left\{
\begin{array}{lll}
w_1 & =Q_p &
\left[2(\lambda_{p}r^2 -6)z(z-1) -2\right]\alpha (z)
+\frac{4}{7}\left[(\lambda_{p}r^2 +6)z(z - \frac{1}{2})(z-1)\right]
\beta (z)\\
w_2 & =Q_p  &
2(1-z)\left[(\lambda_{p}r^2 -6)z +3\right]\alpha (z)
-\frac{4}{7}\left[(\lambda_{p}r^2 +6)z(z - 1)(z-\frac{3}{2})\right]
\beta (z)\\
w_3 &=Q_p&
\left[-2(\lambda_{p}r^2 -6)z(z-1) +3\right]\alpha (z)
-\frac{4}{7}\left[(\lambda_{p}r^2 +6)z(z - \frac{1}{2})(z-1)\right]
\beta (z)
\end{array}
\right.
\ea
with
$\lambda_{p} = (p^2 +3)$ on $S^3$ and $Q_p$ a normalisation constant. 

To fix the normalisation constant $Q_{p}$ we contract the indices in the
coincident limit $z \rightarrow 1$. This yields \cite{Allen}
\begin{equation}
W^{(p)\ ij}_{ij}(\Omega,\Omega)= \sum_{{\cal P}lm}q_{ij}^{(p){\cal P}lm}
(\Omega) q^{(p){\cal P}lm\ ij}(\Omega)^{*}
 =30Q_{p}\alpha (1).
\end{equation}
By integrating over the three-sphere and using the normalisation condition
(\ref{norm}) on the tensor harmonics one obtains
$Q_{p} = -\frac{p^2 +4}{30 \pi^2 \alpha(1)}$.

Notice that (\ref{hyper}) is precisely the hypergeometric 
differential equation, which has a pair of 
independent solutions $\alpha(z)=\  _2F_{1}(3+ip,3-ip,7/2,z)$ and \newline
$\alpha(1-z) =\  _2F_{1}(3+ip,3-ip,7/2,1-z)$.
The former of these solutions is singular at $z=1$, i.e. for coincident points
on the three-sphere, and the latter is singular for opposite points.
The solution for $\beta (z)$ follows from (\ref{bet}) and is given by
\begin{equation}
\beta(z) =\  _2F_{1}(4-ip,4+ip,9/2,z).
\end{equation}
The hypergeometric functions are related by the transformation formula 
(eq.[15.3.6] in \cite{Abram})
\begin{center}
\begin{eqnarray}\label{relat}
_2F_{1}(a,b,c,z)
=\frac{\Gamma (c)\Gamma (c-a-b)}{\Gamma (c-a) \Gamma (c-b)}
 _2F_{1}(a,b,a+b-c,1-z)&&\nonumber\\
 +
\frac{\Gamma (c)\Gamma (a+b-c)}{\Gamma(a)\Gamma(b)}(1-z)^{c-a-b}
\ _2F_{1}(c-a,c-b,c-a-b,1-z).&&
\end{eqnarray}\end{center}
Only for the eigenvalues of the Laplacian on $S^3$, i.e.
$p=in\ (n\geq 3)$,
the term on the second line vanishes for $_2F_{1}(3+ip,3-ip,
7/2,z)$. 
In this case the functions   
are related by $(-1)^{n+1}$ and they are both regular
for any angle on the three-sphere. But since $F(1-z) \rightarrow 1$
for coincident points, we must choose 
$\alpha(1-z)$ in the bitensor
appearing in the Eucliden Green function (\ref{disc}).
This choice also follows from matching the delta function
in the Green equation itself.
In fact, the hypergeometric series
terminates for these parameter values and the hypergeometric functions
reduce to
Gegenbauer's Polynomials $C^{(3)}_{n-3}(1-2z )$.

We conclude that
the above properties required of the bitensor
completely determine its form.
Notice that in terms of the label $p$ we have obtained a
'unified' functional description of the bitensor
$W^{iji'j'}_{(p)}$ on $S^3$ and $H^3$ although its explicit form
is very different in both cases. In fact it is precisely this
which
allowed us in
Section IV to analytically
continue the angular part of the Green function from the Euclidean region
into the open universe.

To perform the continuation we note that the Euclidean geodesic
separation $\mu$ continues to $-i\chi$ where $\chi$ is the
comoving geodesic separation on $H^3$. 
We apply  the relation (\ref{relat}) in an intermediate 
step of
the calculation, the continuation of the bitensor into the
complex $p$-plane. In this step 
the functions $\alpha(z), \beta(z)$
rather than $\alpha(1-z)$ and $\beta (1-z)$ enter.
The hypergeometric functions on $H^3$ are defined by analytic continuation
(eq. 15.3.7 in \cite{Abram}) and
may be expressed
in terms of associated Legendre 
functions as
\ba
\left\{
\begin{array}{ll}
\alpha(z) & =\sqrt{\frac{\pi}{2}}(-\sinh \chi
)^{-5/2} P^{-5/2}_{-1/2 +ip}(-\cosh \chi),
\\
\beta(z) &=\sqrt{\frac{\pi}{2}}(-\sinh \chi
)^{-7/2} P^{-7/2}_{-1/2 +ip}(-\cosh \chi).
\end{array}
\right.
\ea
Using the relation
$-\cosh (\chi) = \cosh (\chi -i\pi)$, 
the Legendre functions may be expressed as
\ba
\left\{
\begin{array}{lll}
P^{-5/2}_{-1/2 +ip}(-\cosh \chi) & = &
\sqrt{\frac{2}{-\pi \sinh \chi}}(1+p^2)^{-1}(4+p^2)^{-1}\left[
3 \coth \chi \cosh p(\pi +i\chi)\right.\\
& & \left. -\frac{i\sinh p(i\chi +\pi)}{2p}\left( (2-p^2)
(1+\coth^2 \chi) +(4+p^2)\mathrm{cosech}^2 \chi \right) \right]\\
P^{-7/2}_{-1/2 +ip}(-\cosh \chi)& = &
\sqrt{\frac{2}{-\pi \sinh \chi}}(1+p^2)^{-1}(4+p^2)^{-1}(9+p^2)^{-1}\times\\
& & \left[
\cosh p(\pi +i\chi)(p^2  -11 - 15 \mathrm{cosech}^2 \chi)\right.\\
& & \left.
-6\frac{i\sinh p(i\chi +\pi)}{p}\left( (1-p^2)\coth^3 \chi
+(p^2+\frac{3}{2})\coth \chi \ \mathrm{cosech}^2 \chi \right) \right]
\end{array}
\right.
\ea
The factors $e^{\pm p\pi}$ in these expressions combine
with similar factors from the continuation of the conformal 
spatial coordinate $X$ to produce our final result (\ref{lorcorf}).
The coefficient functions of the bitensor 
$W_{\ iji'j'}^{L (p)}(\chi)$
in our final result (\ref{lorcorf})
for the tensor correlator are
\ba
\left\{
\begin{array}{lll}
w_1 & = &
\frac{\mathrm{cosech^5 \chi}}{4\pi^2 (p^2 +1)}
\left[\frac{\sin p\chi}{p}(3+(p^2 +4)\sinh^2\chi - 
p^2 (p^2 +1) \sinh^4 \chi)
\right.\\
& & \ \ \ \ \ \ \ \ \ \left. 
+\cos p\chi (3/2 + (p^2 +1) \sinh^2 \chi )\sinh 2\chi)\right]\\
w_2 & = &
\frac{\mathrm{cosech^5 \chi}}{4\pi^2 (p^2 +1)}
\left[\frac{\sin p\chi}{p}(3-12\cosh \chi -3p^2(1-2\cosh \chi )
\sinh^2\chi\right.\\
& & \ \ \ \ \ \ \ \ \ \left. + p^2 (p^2 +1) \sinh^4 \chi) 
-\cos p\chi (-12+3 \cosh \chi \right.\\
& & \ \ \ \ \ \ \ \ \ \left.+2(p^2 -2) \sinh^2 \chi -2(p^2 +1)\cosh \chi 
\sinh^2 \chi) \sinh \chi\right]\\
w_3 & = &
\frac{\mathrm{cosech^5 \chi}}{4\pi^2 (p^2 +1)}
\left[\frac{\sin p\chi}{p}(3 -3p^2\sinh^2\chi +p^2 (p^2 +1) \sinh^4 \chi)\right.\\
& & \ \ \ \ \ \ \ \ \ \left. 
-\cos p\chi (-3/2 + (p^2 +1) \sinh^2 \chi )\sinh 2\chi)\right]\\
\end{array}
\right.
\ea
As mentioned before, for $\chi \rightarrow 0$ these functions converge and
they exponentially decay at large geodesic distances. We also
mention that in this form one should take the normalisation 
factor $Q_p$ to be positive, as explained in the text. 

Finally, let us mention that 
the scalar Green function \cite{Gratton}
may also be described in terms of hypergeometric functions.
In terms of the variable $z$, the equation for
the angular part $C_{p}(\mu)$
of the scalar Euclidean Green function 
(eq.(35) in \cite{Gratton}) reads
\begin{equation}
z(1-z)\frac{d^2 C_{p}(z)}{d^2z} + \left[ \frac{3}{2} -3z\right]
\frac{dC_{p}(z)}{dz}=(p^2 +1)C_{p}(z).
\end{equation}
If we express the Green function as an
infinite sum (eq. (38) in \cite{Gratton}), the appropriate 
solution regular at $\mu=0$ and $\mu=\pi$ is
\begin{equation}
C_{p}(z) = Q_{p}F(1+ip,1-ip,3/2,1-z)
=
\frac{Q_{p}\sinh p\mu}{p\sin \mu}
.\end{equation} 
As for the tensor correlator,
the normalisation constant $Q_{p}$ is 
determined by the normalisation of the scalar harmonics on $S^3$.
However, because of the extra 
factor $(\Delta_{3} +3)$ in the
scalar Green equation (eq.(35) in \cite{Gratton}), we must also divide
by $4+p^2$ in this case.
This reproduces precisely the angular part of the scalar
Green function (eq.(38) in \cite{Gratton}).

When expressing the Euclidean Green function as an integral, we 
continue $C_{p}(z)$ into the complex $p$-plane, and again need
to express it in terms of the hypergeometric function 
regular at $z=0$. We re-express
$F(1+ip,1-ip,3/2,1-z)$ using the 
relation (\ref{relat}) and obtain
\begin{equation}
\coth p\pi\frac{\sinh p \mu}{p\sin \mu} =
\frac{\sinh p(\pi -\mu)}{p\sinh p\pi \sin \mu}+
\frac{ \cosh p\mu}{p \sin \mu}
.\end{equation}
The factor $\coth p\pi$ is needed in converting 
the sum into a contour integral.
The first term is regular for opposite points and leads 
exactly to the angular
part of the Lorentzian correlator (eq.(46) in \cite{Gratton})
in the same way as described above for tensor fluctuations. 
The second term is a bit more subtle.
Its analogue in the tensor correlator 
did not contribute to the contour integral
because it had no poles within the contour. 
However, in the scalar we need to take into account the
extra normalisation factor $\frac{1}{p^2 +4}$ which has a pole at $p=2i$.
This is the underlying reason for the presence of the extra term
in the integral representation of the scalar Euclidean Green function
(2nd term in eq.(37) in \cite{Gratton}). As explained in
\cite{Gratton}), the $(\pi -\Omega)$ factor in front 
of arises from matching the delta function in the Green equation (eq.(35)
in \cite{Gratton}, which unlike the tensor Green equation is
fourth order in derivatives. 
This is also the we had to
include the extra factor $\frac{1}{p^2 +4}$. Nevertheless
it is clear that the scalar and tensor cases are
very closely parallel.

\end{document}